%% file: main.tex
\documentclass{article}

\usepackage[accepted]{icml2023}

\usepackage{amsmath}
\usepackage{amssymb}
\usepackage{mathtools}
\usepackage{amsthm}

\usepackage{graphicx}
\usepackage{booktabs} 
\usepackage{tikz}
\usetikzlibrary{calc, shapes}
\usetikzlibrary{decorations.text}
\usetikzlibrary{arrows.meta}
\usepackage{multirow}
\usepackage{tabularx}
\usepackage{siunitx,mwe}
\usepackage{microtype}

\usepackage{xspace}
\usepackage{hyperref}
\usepackage{url}
\usepackage{subcaption}
\usepackage{tablefootnote}
\usepackage{pgfplots, pgfplotstable}
\pgfplotsset{compat=1.17}
\usetikzlibrary{calc, shapes}
\usepackage[capitalize,noabbrev]{cleveref}
\theoremstyle{plain}

\theoremstyle{definition}

\theoremstyle{remark}

\usepackage[textsize=tiny]{todonotes}

\makeatletter
\DeclareRobustCommand\onedot{\futurelet\@let@token\@onedot}
\def\@onedot{\ifx\@let@token.\else.\null\fi\xspace}

\def\etc{\emph{etc}\onedot} 
 
\def\etal{\emph{et al}\onedot}
\makeatother

\begin{document}

\twocolumn[
\icmltitle{PITS: Variational Pitch Inference Without Fundamental Frequency for End-to-End Pitch-Controllable TTS}


\begin{icmlauthorlist}
\icmlauthor{Junhyeok Lee}{MAUM}
\icmlauthor{Wonbin Jung}{MAUM,KAIST}
\icmlauthor{Hyunjae Cho}{MAUM,SNU}
\icmlauthor{Jaeyeon Kim}{MAUM,SNU}
\icmlauthor{Jaehwan Kim}{MAUM,SNU}
\end{icmlauthorlist}

\icmlaffiliation{MAUM}{maum.ai Inc.}
\icmlaffiliation{SNU}{Seoul National University (SNU)}
\icmlaffiliation{KAIST}{Korea Advanced Institute of Science and Technology (KAIST)}

\icmlcorrespondingauthor{Junhyeok Lee}{jun3518@icloud.com}

\icmlkeywords{text-to-speech, speech synthesis, pitch modeling, variational inference, GAN}
\vskip 0.3in]

\printAffiliationsAndNotice{}

\begin{abstract} 
Previous pitch-controllable text-to-speech (TTS) models rely on directly modeling fundamental frequency, leading to low variance in synthesized speech. To address this issue, we propose PITS, an end-to-end pitch-controllable TTS model that utilizes variational inference to model pitch. Based on VITS, PITS incorporates the Yingram encoder, the Yingram decoder, and adversarial training of pitch-shifted synthesis to achieve pitch-controllability. Experiments demonstrate that PITS generates high-quality speech that is indistinguishable from ground truth speech and has high pitch-controllability without quality degradation. Code, audio samples, and demo are available at https://github.com/anonymous-pits/pits.
\end{abstract}

\setlength{\textfloatsep}{12pt}%

\input{paper/introduction}

\input{paper/method}

\input{paper/experiments}

\input{paper/results}

\input{paper/discussion}


\bibliographystyle{icml2023}
\bibliography{main}

\newpage
\appendix
\input{paper/appendix}

\end{document}

%% file: paper/introduction.tex
\section{Introduction}
Text-to-speech (TTS) is a generative task that synthesizes human-like speech waveforms from a given text input.
Since it is a one-to-many mapping problem, there are some challenges to model diverse duration, pitch, and \etc in speeches.
Recent deep neural network-based TTS models \cite{tacotron, tacotron2,vits,eats,fastspeech2,trinitts,periodvits} have significantly increased their synthesis quality.
Among them, VITS \cite{vits} is the most successful end-to-end (E2E) TTS, which is modeling the one-to-many nature of TTS by adopting variational auto-encoder (VAE) \cite{vae} and normalizing flow \cite{residualflows} to provide diverse pitches and rhythms.
Especially in rhythms, J. Kim \etal \cite{vits} propose a stochastic duration predictor and monotonic alignment search (MAS) that finds the maximum evidence lower bound (ELBO) path between the latent variables from the short-time Fourier transform (STFT) encoder and the text encoder to train a model without annotated duration labels.


On the other hand, there have been several efforts to model and control the pitch of synthesized speech. 
FastSpeech 2 \cite{fastspeech2} adopts a variance adaptor with a pitch predictor that predicts fundamental frequency ($f_0$) at the frame-level to provide pitch information in the speech.
FastPitch \cite{fastpitch} improves upon FastSpeech 2 by proposing character-level pitch prediction,
and FastPitchFormant \cite{fastpitchformant} builds on these previous methods by incorporating pitch information to model speech according to the source-filter theory \cite{source_filter_theory}.
These models are deterministic in pitch predictions and only rely on the variance in dropout.
VarianceFlow \cite{varianceflow} applies normalizing flow for variance, but it also has low variance due to directly modeling $f_0$.
Although text-to-pitch is a one-to-many mapping, these pitch-controllable TTS models tend to have low pitch variance while they are directly modeling ground truth $f_0$.

There are also some variants that have been developed to adopt pitch predictors in VITS.
VISinger \cite{visinger}, a singing voice synthesizer, incorporates frame prior networks and $f_0$ predictor for enhanced acoustic variation and more natural singing. 
However, the use of frame prior networks requires the duration predictor to be trained with annotated labels without the use of MAS.
Period VITS \cite{periodvits}, builds on the modifications introduced in VISinger and adds a periodicity generator. 
These models have limitations in that it requires external annotated duration labels and pitch contours which are expected to reduce the diversity of speech synthesis.

Since entangled pitch information, especially $f_0$, in linguistic information degrades the quality of speech synthesis, 
there have been several efforts to disentangle them in voice conversion \cite{speechresynthesis,nansy,assemvc,vqmivc}. 
A. Polyak \etal \cite{speechresynthesis} attempt to separate these two factors using vector quantization (VQ) models, HuBERT \cite{hubert} and VQ-VAE \cite{vqvae}.
H. Choi \etal \cite{nansy} use wav2vec 2.0 \cite{wav2vec2} for language information and introduced Yingram, an acoustic feature inspired by YIN algorithm \cite{yin}, that captures pitch information including harmonics. 
Yingram is designed to address the limitations of $f_0$, which is not well-defined in some cases \cite{gcidetect}.
This feature enables the training of pitch-controllable models in a fully self-supervised manner.
In addition, the Yingram-based model shows better preference than the  $f_0$-based model \cite{nansy}.

To overcome the limitations of previous pitch-controllable TTS models, which have low variance and rely on external $f_0$ extractors and labeled phoneme durations, we propose \textit{PITS}, variational Pitch Inference for Text-to-Speech, an E2E model that incorporates variational pitch inference without $f_0$.
PITS models pitch information by using conditional VAE and normalizing flow for the Yingram, instead of directly modeling $f_0$.
Additionally, PITS utilizes two posterior encoders, the STFT encoder, and an additional Yingram encoder, unlike the original VITS, which only utilizes a single posterior encoder.
We also introduce several new features to PITS to improve pitch-controllability, including the Yingram decoder and pitch-shifted waveform synthesis during training.
To support these new features, we propose additional loss terms, including Yingram reconstruction loss, Yingram decoding loss, and adversarial pitch-shifted loss.
Through our proposed method, PITS can generate speech with high naturalness, almost indistinguishable from recorded speech.
In addition, PITS has pitch-controllability that could shift its relative pitch without quality degradation.

%% file: paper/method.tex
\section{PITS}
\begin{figure*}[th!]
  \centering
  \subcaptionbox{PITS training procedure}%
  [.513\textwidth]{\includegraphics[height=.34\textwidth]{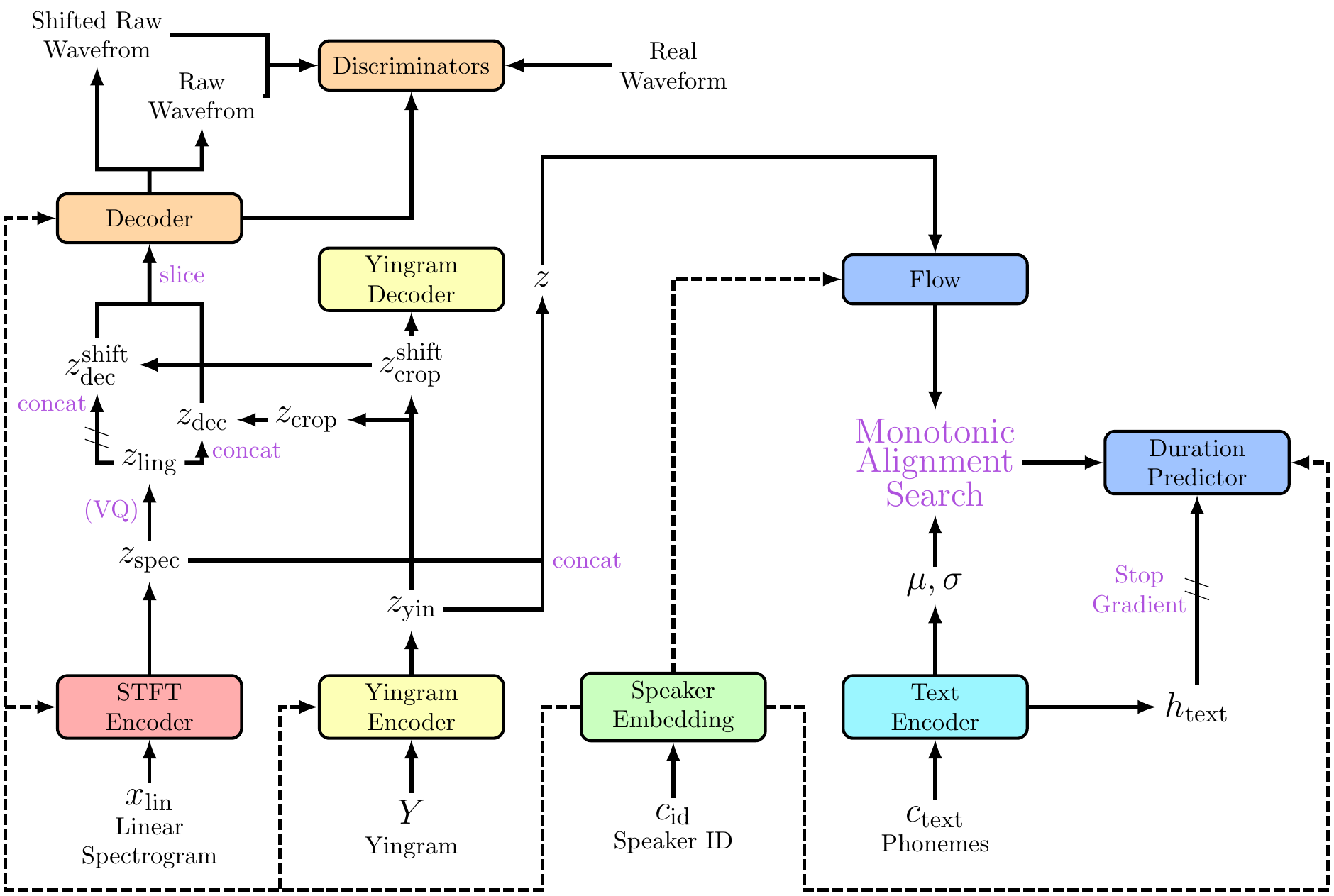}}
  \vrule
  \subcaptionbox{PITS inference procedure}%
  [.477\textwidth]{\includegraphics[height=.34\textwidth]{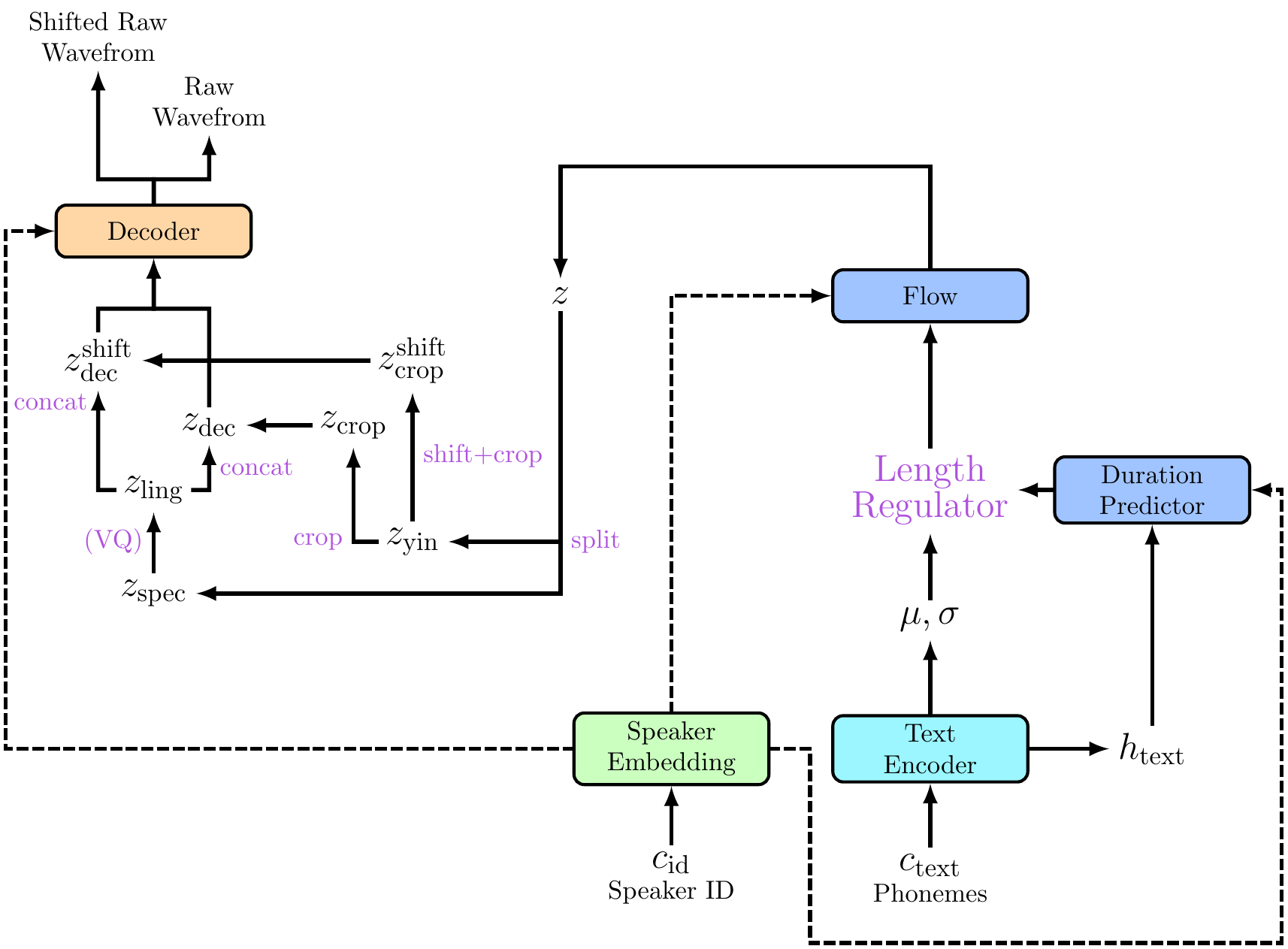}}
      \vspace{-0.75\baselineskip}
\caption{Overall system diagram of (a) training procedure and (b) inference procedure.}
\vspace{-0.75\baselineskip}
\label{fig:fig}
\end{figure*}

Based on the VITS generator \cite{vits}, we add several modifications to build a pitch-controllable model. 
Different from VITS, PITS has Yingram encoder, Yingram decoder, Q-VAE, and synthesizing pitch-shifted waveforms during training. 
Figure \ref{fig:fig} is an overall illustration of PITS during training and inference.

\subsection{Yingram Encoder}
We propose an additional posterior encoder, the Yingram encoder, to model pitch information including harmonics. 
Similar to previous works aimed at disentangling linguistic information and pitch information \cite{speechresynthesis,nansy}, our proposed model PITS uses both the STFT encoder and the Yingram encoder to encode linguistic information and pitch information, respectively.
The architecture of the Yingram encoder is identical to the STFT encoder without the number of channels.
We set the number of input and output channels to be the same as 80 channels.

To achieve pitch-controllability, we crop the latent variables from the Yingram encoder, instead of cropping the Yingram as done in NANSY \cite{nansy}.
While voice conversion tasks have access to a source speech to calculate the Yingram, TTS does not have access to any source speech during inference. 
Therefore, we crop the middle (16th - 65th channels) of the latent variables in the channel dimension, similar to cropping middle-frequency bins of Yingram in NANSY.
We set the scope of the latent variables for the decoder as 50 channels, while we are utilizing 80 channels Yingram.
To start, the latent variables from the Yingram encoder, $z_\mathrm{yin}$, are cropped to default scope (16th - 65th channels) as $z_\mathrm{crop}$.
Additionally, scope-shift is determined by the randomly sampled integer $s$, which represents scope-shift, from a uniform distribution between $[-15,15]$, and crops the scope-shifted (($s$+16)th - ($s$+65)th channels) latent variables $z_\mathrm{crop}^\mathrm{shift}$.
The cropped latent variables from the Yingram encoder are concatenated with the latent variables from the STFT encoder and concatenated latent variables are fed to the decoder.

For the flow and MAS, we concatenate the variational latent variables from the Yingram encoder and the STFT encoder and feed them to the normalizing flow.
The ELBO or Kullback–Leibler (KL) divergence calculations are also the same as in the original VITS without concatenation.

The Yingram encoder can be employed for voice conversion (VC) by extracting pitch information from the source samples. VC details are available in Appendix \ref{appx:vc}.

\subsection{Yingram Decoder}
Building pitch-controllability by shifting the scope of latent variables requires the Yingram encoder to be translation equivariant in the channel direction.
However, we found that outside of the default scope  was not trained and remained as Gaussian. 
To enforce this translation equivariance, we design the Yingram decoder to reconstruct shifted scopes of Yingram  from shifted latent variables.
The Yingram decoder then reconstructs the Yingram of ($s$+16)th - ($s$+65)th channels from the scope-shifted latent variables $z_\mathrm{crop}^\mathrm{shift}$.
$L^1$ norm is used to measure this reconstruction, the Yingram decoding loss $\mathcal{L}_{yd}$ is represented as:
\begin{equation}
    \mathcal{L}_{yd} = \lambda_\mathrm{yin} \mathbb{E}\left[ \left\Vert Y_\mathrm{crop}^\mathrm{shift} - \mathrm{Dec}_\mathrm{yin} \left(z_\mathrm{crop}^\mathrm{shift} \right) \right\Vert_1 \right],
\end{equation}
where $Y$ is the original Yingram, $Y_\mathrm{crop}$ is the cropped Yingram with default scope, $Y_\mathrm{crop}^\mathrm{shift}$ is the cropped Yingram with scope-shift, $\mathrm{Dec}_\mathrm{yin}$ is the Yingram decoder, and $\lambda_\mathrm{yin}$ is a multiplier for yin reconstruction loss and set to 45.

\subsection{Q-VAE}
We try Q-VAE \cite{QFVAE,stylelabelfree} for the output of the STFT encoder $z_\mathrm{spec}$ to disentangle the outputs of the Yingram encoder and the STFT encoder.
We anticipate that quantizing the latent variables from the STFT encoder will effectively separate pitch information from linguistic features, while pitch information is typically represented as a continuous variable.
However, previous TTS models including VQ \cite{QFVAE,vqtts,wavthroughvec} have a limitation in that they are multiple-stage models. 
To remove multiple stages including VQ extractor and vocoder training, we apply Q-VAE to the VITS training scheme \cite{vits} by quantizing the output latent variable of the STFT encoder $z_\mathrm{spec}$.
With this implementation, the VQ extractor is trained along the waveform decoder, flow, text encoder, and duration predictor.

During applying Q-VAE \cite{QFVAE,stylelabelfree}, we observed that codebook loss $\mathcal{L}_\mathrm{vq}$ and commit loss $\mathcal{L}_\mathrm{commit}$ diverge.
It seems that our scheme is different from normal VQ-VAE since $z_\mathrm{spec}$ gets multiple gradients  from flow and codebooks.
To prevent the diverging loss, we scale the gradient of $z_\mathrm{crop}^\mathrm{shift}$ from codebooks to $1/\lambda_\mathrm{mel}$, where $\lambda_\mathrm{mel}$ is a multiplier for mel-spectrogram reconstruction loss and set to 45. 
We use codebooks of size 160 and apply two codebooks as multiple codebooks in previous quantization modules \cite{hubert,wav2vec2}.


\subsection{Pitch-Shifted Waveform Synthesis}
The cropped latent variables from the Yingram encoder, $z_\mathrm{crop}$ and $z_\mathrm{crop}^\mathrm{shift}$, are concatenated with the latent variables from the STFT encoder, $z_\mathrm{vq}$ or $z_\mathrm{spec}$, depending on whether quantization is applied, to form $z_\mathrm{dec}$ and $z_\mathrm{dec}^\mathrm{shift}$ for input to the decoder.
For brevity, $z_\mathrm{vq}$ or $z_\mathrm{spec}$ is replaced to $z_\mathrm{ling}$.
In the case of scope-shifted variables, $z_\mathrm{crop}^\mathrm{shift}$ is concatenated with $z_\mathrm{ling}$ after applying gradient stop to disentangle pitch information from it.

To construct a stable pitch-controllable model, PITS synthesizes pitch-shifted waveforms during training.
While the default synthesized raw waveforms are trained to minimize mel-spectrogram loss to reconstruction loss,
pitch-shifted waveforms do not have a corresponding ground-truth mel-spectrogram as the default synthesized signals do.
To ensure that pitch-shifted synthesized speeches are indeed ``pitch-shifted" from the default synthesized speeches,
we propose two additional loss terms, the Yingram reconstruction loss, and the adversarial pitch-shifted loss.

First, since the ground truth mel-spectrogram of pitch-shifted speech is not available, we apply the Yingram reconstruction loss to both of normal synthesized signals and pitch-shifted synthesized signals for reconstruction loss. 
As YIN \cite{yin} and Yingram \cite{nansy} are autocorrelation-based algorithms, they can be made differentiable.
However, unlike the Yingram decoding, the ideal Yingram of the pitch-shifted speech could not be perfectly translation equivariant due to the presence of linguistic information.
Thus, we apply a negative exponential to the cropped Yingrams, as values close to zero are more critical for representing harmonics.
Yingram reconstruction loss $\mathcal{L}_\mathrm{yin}$ is defined as:
\begin{align}
    \mathcal{L}_\mathrm{yin} = &  \lambda_\mathrm{yin}\mathbb{E}
    \left[
        \left\Vert 
            e^{ \left( - Y_\mathrm{crop} \right)}
            - e^ {\left( - \mathcal{Y}_\mathrm{crop}\left( G \left(z_\mathrm{dec} \right) \right) \right) }
        \right\Vert_1 
    \right. \nonumber \\
     &\left.+\left\Vert
            e^{ \left( - Y_\mathrm{crop}^\mathrm{shift} \right)} 
            - e^ {\left( - \mathcal{Y}_\mathrm{crop}\left( G \left(z_\mathrm{dec}^\mathrm{shift} \right) \right) \right) }
     \right\Vert_1 \right],
\end{align}
where $\mathcal{Y}_\mathrm{crop}$ is the Yingram crop function and $G$ is the waveform decoder. In addition, while we apply additional Yingram reconstruction loss, we modify the notation of mel-spectrogram reconstruction loss from $\mathcal{L}_\mathrm{recon}$ to $\mathcal{L}_\mathrm{mel}$.

Second, we encountered quality degradation in the pitch-shifted synthesized speech by the model that was trained using only $\mathcal{L}_\mathrm{yin}$.
To prevent this degradation, we also feed pitch-shifted synthesized speech to the discriminators for adversarial feedback to pitch-shifted speeches.
Since waveform synthesis GANs \cite{vits,hifigan,melgan,avocodo} get a paired input, we also pair the pitch-shifted waveforms with their original real waveforms.
The time-aligned linguistic features of the pitch-shifted waveforms and real waveforms are identical, despite the difference in pitch, which makes the paired adversarial training between them reliable.
Identical to waveform synthesis GANs, we apply the least-square loss \cite{lsgan} $\mathcal{L}_\mathrm{adv}^\mathrm{shift}(G)$ for the GAN and feature matching loss \cite{featurematching} $\mathcal{L}_\mathrm{fm}^\mathrm{shift}(G)$ for the generator.

For the discriminators, we apply CoMBD and SBD from Avocodo \cite{avocodo} instead of MPD \cite{hifigan}. 
We expect that CoMBD's collaborative structure, with multi-scale and hierarchical inputs, could enhance speech quality.
To improve the pitch-shifted synthesis, we also feed hierarchical outputs of pitch-shifted synthesis to the CoMBD.
To prevent periodicity artifacts, we apply PhaseAug \cite{phaseaug} during training.
We do not apply PhaseAug to hierarchical outputs from the generator.


\subsection{Total Loss}
With our proposed method, the total loss for training PITS can be represented as follows:
\begin{align} \label{eq:loss_total}
    \mathcal{L}_\mathrm{total} =&\mathcal{L}_\mathrm{mel} +\mathcal{L}_\mathrm{KL} +\mathcal{L}_\mathrm{dur} + \mathcal{L}_\mathrm{adv}(G)+ \mathcal{L}_\mathrm{fm}(G) \nonumber  \\
    &+\mathcal{L}_\mathrm{yin} + \left( \mathcal{L}_\mathrm{adv}^\mathrm{shift}(G) + \mathcal{L}_\mathrm{fm}^\mathrm{shift}(G) \right)\nonumber\\ 
    & + \left(\mathcal{L}_\mathrm{yd} \right)+ \left( \mathcal{L}_\mathrm{vq}+\mathcal{L}_\mathrm{commit} \right),
\end{align}
where $\mathcal{L}_\mathrm{mel}$, $\mathcal{L}_\mathrm{KL}$, $\mathcal{L}_\mathrm{dur}$, $\mathcal{L}_\mathrm{adv}$, and $\mathcal{L}_\mathrm{fm}$ are originated from loss of VITS \cite{vits}. Terms enclosed in parentheses can be omitted in the model architecture search.

%% file: paper/experiments.tex
\input{paper/table_pitch}

\section{Experiments}
\subsection{Datasets}
We conduct experiments with the VCTK dataset \cite{vctk}, which includes 44 hours of recorded speech by 108 speakers with text labels.
We preprocess the dataset by resampling all audio files to 22.05 kHz sampling rate data and transcribing all text into ARPABET without stress using the internal grapheme-to-phoneme model. 
The dataset is then split into training, validation, and test sets, with 43,225, 324, and 324 files, respectively.

\subsection{Model Setups}
To determine the optimal architecture, we perform an architecture search with four combinations.
Although the Yingram encoder and $\mathcal{L}_\mathrm{yin}$ are necessary for constructing a pitch-controllable model, we do not include them in the architecture search.
The four variations of PITS are (A+D+Q), (A+D), (A+Q), and (D+Q), where A, D, and Q respectively denote adversarial loss for pitch-shifted signal, Yingram decoding loss, and Q-VAE.
To evaluate the quality of pitch-controllability, we generate the pitch-shifted speeches by shifting the scope of normal speech, corresponding to a scope-shift of 0, with scope-shift values of [8,6,4,2,-2,-4,-6,-8], which correspond to pitch-shift values of [-4,-3,-2,-1,+1,+2,+3,+4] semitones, respectively.   

We compare normal speech synthesis of PITS with previous state-of-the-art TTS models, VITS \cite{vits} and FastSpeech 2 + HiFi-GAN (FS2) \cite{fastspeech2,hifigan}.
We trained VITS with ARPABET without stress by modifying the official implementation. 
For FS2, we adopted an unofficial modified implementation with unsupervised duration \cite{onealignment} and its pre-trained model\footnote{The pre-trained modified FastSpeech 2 model was trained on the VCTK and ARPABET datasets with stress.} \cite{comp_tts}. HiFi-GAN was trained by ourselves using the official implementation.

All models, except for FS2, were trained for 3,000 epochs from scratch on four V100 GPUs.
Automatic mixed precision was applied during training.
We set the batch size to 48 for each GPU.
Each training took between 14 to 18 days.

\subsection{Implementation Details}
The flow model consists of 192 channels, the STFT encoder consists of 112 channels, and the decoder gets 162 channels of input.
Aside from these modifications, the remaining details of the architecture are identical to the original VITS setup \cite{vits}.
Our Yingram setup is different from that used in H. Choi \etal \cite{nansy}.
In our setup, 24 notes represent an octave for fine pitch modeling and note 69 is fixed at 440 Hz. 
We generate Yingram with 80 channels from notes ranging between -5 to 74, which corresponds to 30.8 to 508 Hz.

\subsection{Evaluation}
To compare the models, we generate speeches from the models only with the test split's text and speaker identity.
We conduct a subjective evaluation using 5-scale mean opinion score (MOS).
Each of MOS is performed with 324 samples and 1,620 ratings, with 95\% confidence interval (CI).
To quantitatively evaluate the quality of TTS models, we measure equal error rate (EER) and character error rate (CER) to assess their speaker similarity and speech intelligibility, respectively.
EER is measured by official pre-trained RawNet3 \cite{rawnet3}.
CER is calculated by a pre-trained open-source model, Silero xlarge \cite{silero}.
To measure EER and CER, we resample all audio signals to 16 kHz.

To evaluate the quantitative performance of pitch-shift, we use the shifted part of the Yingram reconstruction loss as
$\scriptsize{\mathcal{L}_\mathrm{yin}^\mathrm{shift}= \lambda_\mathrm{yin}  \mathbb{E}
    \left[\left\Vert
            e^{ \left( - \mathcal{Y}_\mathrm{crop}^\mathrm{shift}(z_\mathrm{dec}) \right)} 
            - e^ {\left( - \mathcal{Y}_\mathrm{crop}\left( G \left(z_\mathrm{dec}^\mathrm{shift} \right) \right) \right) }
     \right\Vert_1 \right]. \nonumber}$
Unlike the loss calculation during training, where the ground truth is available, we compute the difference between the Yingram of normal synthesized speech and scope-shifted speech to obtain a $\mathcal{L}_\mathrm{yin}^\mathrm{shift}$.

%% file: paper/table_pitch.tex
\setcounter{table}{1}
\begin{table*}[th!]
  \caption{
  The evaluation results of pitch-shifted synthesized speeches. The best value in each pitch-shift is shown in bold. In cases where the standard deviation of two values overlaps, The values are paired in bold. While the (D+Q) model, which is expected to overfitting in ${\mathcal{L}_\mathrm{yin}^\mathrm{shift}}$, yields the best results in ${\mathcal{L}_\mathrm{yin}^\mathrm{shift}}$, the second-best value is also highlighted in bold.
  }
  \vspace{-0.6\baselineskip}
  \label{tab:ps}
  \centering
  \resizebox{0.99\linewidth}{!}{%
  \begin{tabular}{l| cccc | cccc | cccc | cccc}
    \toprule
     & \multicolumn{4}{c|}{\textbf{MOS} $\pm$ \textbf{CI} ($\uparrow$)} & \multicolumn{4}{c|}{\textbf{CER (\%)} ($\downarrow$)} & \multicolumn{4}{c|}{\textbf{EER (\%)} ($\downarrow$)} & \multicolumn{4}{c}{${\mathcal{L}_\mathrm{yin}^\mathrm{shift}}$ ($\downarrow$)} \\
     \midrule
    \phantom{-}$s$ &
    A+D+Q &  \textbf{A+D} &  A+Q & D+Q & 
    A+D+Q &  \textbf{A+D} &  A+Q & D+Q &  
    A+D+Q &  \textbf{A+D} &  A+Q & D+Q & 
    A+D+Q &  \textbf{A+D} &  A+Q & \textbf{D+Q}  \\
    \midrule
    \phantom{-}8    
    & $3.39$\scriptsize{$\pm$0.05}  & $\bf{3.87}$\scriptsize{$\pm$0.04} & $3.72$\scriptsize{$\pm$0.05} & $2.39$\scriptsize{$\pm$0.06}
    & $57.1$ & $\bf{5.76}$ & $20.9$ & $52.0$
    & $22.2\phantom{0}$ & $\bf{7.10\phantom{0}}$ & $10.3\phantom{0}$ & $38.6\phantom{0}$ 
    & $1.34$ & $\bf{1.01\phantom{0}}$ & $1.08$ & $\bf{0.927}$ \\
    \phantom{-}6    
    &$3.41$\scriptsize{$\pm$0.05}  & $\bf{3.94}$\scriptsize{$\pm$0.04} & $3.77$\scriptsize{$\pm$0.04} & $2.48$\scriptsize{$\pm$0.06} 
    & $56.4$ & $\bf{5.00}$ & $19.7$ & $51.5$ 
    & $14.0\phantom{0}$ & $\bf{3.86\phantom{0}}$ & $\phantom{0}7.25$ & $35.8\phantom{0}$ 
    & $1.34$ & $\bf{0.959}$ & $1.04$ & $\bf{0.934}$ \\
    \phantom{-}4    
    & $3.45$\scriptsize{$\pm$0.05}  & $\bf{3.92}$\scriptsize{$\pm$0.04} & $3.80$\scriptsize{$\pm$0.04} & $2.68$\scriptsize{$\pm$0.06} 
    & $58.8$ & $\bf{3.98}$ & $18.3$ & $51.3$
    & $10.2\phantom{0}$ & $\bf{2.62\phantom{0}}$ & $\phantom{0}5.25$ & $32.4\phantom{0}$ 
    & $1.32$ & $\bf{0.982}$ & $1.11$ & $\bf{0.938}$ \\
    \phantom{-}2    
    & $3.49$\scriptsize{$\pm$0.05}  & $\bf{4.01}$\scriptsize{$\pm$0.04} & $3.85$\scriptsize{$\pm$0.04} & $2.73$\scriptsize{$\pm$0.06}
    & $55.4$ & $\bf{3.64}$ & $17.5$ & $50.8$
    & $\phantom{0}5.40$ & $\bf{1.85\phantom{0}}$ & $\phantom{0}3.09$ & $26.5\phantom{0}$ 
    & $1.52$ & $\bf{1.15\phantom{0}}$ & $1.28$ & $\bf{0.927}$ \\
    \midrule
    \phantom{-}0    
    & $3.70$\scriptsize{$\pm$0.05}  & $\bf{4.01}$\scriptsize{$\pm$0.04} & $3.90$\scriptsize{$\pm$0.04} & $3.65$\scriptsize{$\pm$0.05} 
    & $42.3$ & $\bf{3.27}$ & $18.8$ & $49.6$
    & $\phantom{0}2.16$ & $\bf{0.926}$& $\phantom{0}1.39$ & $\phantom{0}2.16$ 
    &   $-$   &   $-$   &    $-$&$-$  \\
    \midrule
    -2    
    & $3.48$\scriptsize{$\pm$0.05}  & $\bf{3.94}$\scriptsize{$\pm$0.04} & $3.79$\scriptsize{$\pm$0.04} & $2.57$\scriptsize{$\pm$0.06} 
    & $56.2$ & $\bf{3.66}$ & $18.1$ & $52.4$
    & $\phantom{0}6.17$ & $\bf{2.16\phantom{0}}$ & $\phantom{0}4.32$ & $22.7\phantom{0}$ 
    & $1.53$ & $\bf{1.20\phantom{0}}$ & $1.29$ & $\bf{0.957}$ \\
    -4    
    & $3.51$\scriptsize{$\pm$0.05}  & $\bf{3.84}$\scriptsize{$\pm$0.04} & $\bf{3.78}$\scriptsize{$\pm$0.04} & $2.52$\scriptsize{$\pm$0.06} 
    & $58.0$ & $\bf{4.17}$ & $19.2$ & $53.0$
    & $12.7\phantom{0}$ & $\bf{3.40\phantom{0}}$ & $\phantom{0}6.63$ & $32.4\phantom{0}$ 
    & $1.34$ & $\bf{1.05\phantom{0}}$ & $1.16$ & $\bf{1.01\phantom{0}}$ \\
    -6    
    & $3.51$\scriptsize{$\pm$0.05}  & $\bf{3.79}$\scriptsize{$\pm$0.05} & $\bf{3.73}$\scriptsize{$\pm$0.05} & $2.53$\scriptsize{$\pm$0.06} 
    & $58.1$ & $\bf{4.83}$ & $20.7$ & $54.3$ 
    & $20.2\phantom{0}$ & $\bf{5.71\phantom{0}}$ & $12.0\phantom{0}$ & $37.5\phantom{0}$ 
    & $1.36$ & $\bf{1.06\phantom{0}}$ & $1.12$ & $\bf{1.05\phantom{0}}$ \\
    -8    
    & $3.50$\scriptsize{$\pm$0.05}  & $\bf{3.78}$\scriptsize{$\pm$0.05} & $3.64$\scriptsize{$\pm$0.05} & $2.50$\scriptsize{$\pm$0.06} 
    & $58.7$ & $\bf{5.84}$ & $23.0$ & $54.5$
    & $24.4\phantom{0}$ & $\bf{6.94\phantom{0}}$ & $15.1\phantom{0}$ & $40.9\phantom{0}$ 
    & $1.44$ & $\bf{1.10\phantom{0}}$ & $1.18$ & $\bf{1.08\phantom{0}}$\\
    \bottomrule
  \end{tabular}
  }
      \vspace{-0.75\baselineskip}
\end{table*}

%% file: paper/results.tex
\section{Results}
\vspace{-0.5\baselineskip}
Table \ref{tab:result} presents the results of normal speech synthesis.
The evaluation metrics, including MOS, CER, and EER, indicate that PITS (A+D) outperforms VITS and FS2.
In addition, the MOS of PITS (A+D) does not have a statistically significant difference from that of the ground truth samples.
  \vspace{-0.75\baselineskip}
\setcounter{table}{0}
\begin{table}[ht]
  \caption{The evaluation results of models.}
    \vspace{-0.75\baselineskip}
  \label{tab:result}
  \centering
  \resizebox{0.92\linewidth}{!}{%
  \begin{tabular}{l ccc}
    \toprule
    \textbf{Model} & \textbf{MOS} $\pm$ \textbf{CI} ($\uparrow$) & \textbf{CER (\%)} ($\downarrow$)  & \textbf{EER (\%)} ($\downarrow$)\\
    \midrule
    Ground Truth & $ 4.04 \pm 0.04$ & $\phantom{0}2.63$& $1.54\phantom{0}$\\
    \midrule
    VITS & $3.89 \pm 0.04$ & $\phantom{0}4.35$ & $1.08\phantom{0}$\\
    FS2 & $3.84 \pm 0.04$ & $\phantom{0}3.78$ & $2.62\phantom{0} $ \\
    PITS (A+D+Q) & $3.70 \pm 0.05$& $42.3\phantom{0}$ & $2.16\phantom{0}$ \\
    \textbf{PITS (A+D)} & $\bf{4.01 \pm 0.04}$ & $\bf{\phantom{0}3.27}$ & $\bf{0.926}$\\
    PITS (A+Q) & $3.90 \pm 0.04$ & $18.8\phantom{0}$& $1.39\phantom{0}$\\
    PITS (D+Q) & $3.65 \pm0.05$ & $49.6\phantom{0}$ & $2.16\phantom{0}$\\
    \bottomrule
  \end{tabular}
  }
    \vspace{-0.75\baselineskip}
\end{table}

Table \ref{tab:ps} reports the evaluation results of the pitch-shifted speech.
While PITS (A+D) is also the best combination in pitch-shifted synthesis without quality degradation, there are some noticeable points.
The results demonstrate that the Q-VAE leads to a degradation in the quality and intelligibility of synthesized speech, as evidenced by a significant increase in the CER of the Q-VAE-included models.
Moreover, combining the Yingram decoding loss with Q-VAE results in further deterioration in the quality and intelligibility of synthesized speech.
The results for EER indicate that pitch-shift affects the speaker's identity.
Although PITS (A+D) exhibits the smallest EER among all shifts, it increases as the pitch-shift increases.
A comparison between (A+D+Q) and (D+Q) indicates that the adversarial loss for pitch-shift enhances the naturalness of pitch-shifted synthesis. 
However, there is no significant difference in the naturalness of synthesized speech during normal inference. 
In addition, when the A option is not used, we can observe that PITS (D+Q) focuses on reducing ${\mathcal{L}_\mathrm{yin}^\mathrm{shift}}$ without quality improvements.
Without PITS (D+Q), PITS (A+D) also outperforms other models in ${\mathcal{L}_\mathrm{yin}^\mathrm{shift}}$.

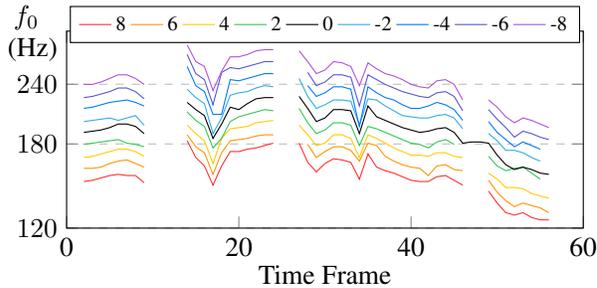
\begin{figure}[ht]
  \centering
  \resizebox{\linewidth}{!}{
  \input{figure/pitch_contour}}
    \vspace{-1.75\baselineskip}
  \caption{Pitch contours of pitch-shifted speeches.}
  \label{fig:pitch_contour}
    \vspace{-0.5\baselineskip}
\end{figure}

\begin{figure}[ht]
  \centering
  \resizebox{\linewidth}{!}{
  \input{figure/pitch_variaiton}}
    \vspace{-1.75\baselineskip}
  \caption{Pitch contours of synthesized speech with identical text, ``How much variation is there?", and identity, p277.}
  \label{fig:pitch_var}
    \vspace{-0.75\baselineskip}
\end{figure}
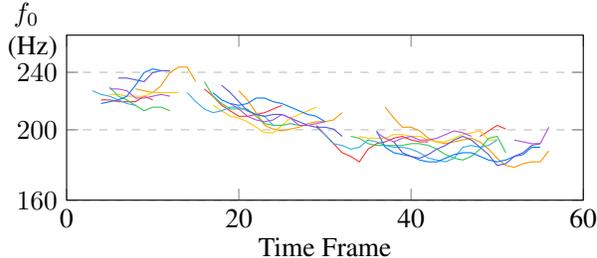

Figure \ref{fig:pitch_contour} depicts the pitch contours of normal and pitch-shifted synthesized speeches, extracted by PYIN \cite{pyin}.
The figure illustrates that the pitch contours are almost parallel in the log scale, which confirms that our proposed method has high pitch-controllability.
Figure \ref{fig:pitch_var} illustrates that PITS could generate speeches with diverse pitches from identical text and speaker identity by modeling pitch information using a variational Yingram encoder.

\vspace{-0.5\baselineskip}
\vspace{-1.\baselineskip}

%% file: figure/pitch_contour.tex
\begin{tikzpicture}

\global\definecolor{ired}{RGB}{255,49,48}
\global\definecolor{iorange}{RGB}{255,149,0}
\global\definecolor{iyellow}{RGB}{255,204,0}
\global\definecolor{igreen}{RGB}{52,199,89}
\global\definecolor{icyan}{RGB}{50,173,230}
\global\definecolor{iblue}{RGB}{0,122,255}
\global\definecolor{iindigo}{RGB}{88,86,214}
\global\definecolor{ipurple}{RGB}{175,82,222}

\begin{axis}[
    unbounded coords=jump,
    xlabel={Time Frame},
    ylabel={$f_0$\phantom{z}\\(Hz)},
    ymode=log,
    log ticks with fixed point,
    width=\linewidth,
    height=\linewidth*0.5,
    xmin=10, xmax=70,
    ymin=120, ymax=310,
    xtick={10,30,50,70},
    xticklabels={0,20,40,60},
    ytick={120,180,240},
    legend image post style={scale=0.695},
    legend style={at={(0.5,+1.125)},anchor=north,nodes={scale=0.75, transform shape}, outer sep=0pt},
    legend columns=9,
    ymajorgrids=true,
    grid style=dashed,
    label style={inner sep=0pt},
    every axis y label/.style={
        at={(0,0.8)},
        anchor=south east,
        align=center,
        outer sep=0pt,
    },
    outer sep=0pt,
]

\addplot[
    color=ired,
    ]
coordinates {
(0,nan)(1,nan)(2,nan)(3,nan)(4,nan)(5,nan)(6,nan)(7,nan)(8,nan)(9,nan)(10,nan)(11,nan)(12,150.19585059339548)(13,151.06592630451914)(14,152.82122781686914)(15,154.59692495035372)(16,155.49249582586248)(17,154.59692495035372)(18,154.59692495035372)(19,149.33078614961454)(20,nan)(21,nan)(22,nan)(23,nan)(24,182.78887791090543)(25,168.58914210463126)(26,161.90845090664698)(27,147.61557577921405)(28,161.90845090664698)(29,173.52918104210445)(30,173.52918104210445)(31,175.5454930018647)(32,176.56241769755178)(33,178.6139741695516)(34,180.68936857950385)(35,nan)(36,nan)(37,180.68936857950385)(38,164.73856077731807)(39,157.29923159850804)(40,163.7897364863335)(41,167.61814009714527)(42,166.65273064969477)(43,164.73856077731807)(44,151.9410423129619)(45,171.53602840047625)(46,160.97592684906726)(47,158.21045677717683)(48,156.39325469067515)(49,153.70651218352927)(50,151.06592630451914)(51,150.19585059339548)(52,150.19585059339548)(53,153.70651218352927)(54,154.59692495035372)(55,151.06592630451914)(56,147.61557577921405)(57,nan)(58,nan)(59,143.41324688128702)(60,135.36407982952485)(61,129.25125509628117)(62,127.76667781088264)(63,129.25125509628117)(64,126.29915234996876)(65,124.84848285663055)(66,124.84848285663055)(67,nan)(68,nan)(69,nan)(70,nan)(71,nan)(72,nan)(73,nan)(74,nan)(75,nan)(76,nan)(77,nan)(78,nan)(79,nan)(80,nan)};
\addlegendentry{8};

\addplot[
    color=iorange,
    ]
coordinates {
(0,nan)(1,nan)(2,nan)(3,nan)(4,nan)(5,nan)(6,nan)(7,nan)(8,nan)(9,nan)(10,nan)(11,nan)(12,160.04877373483913)(13,160.04877373483913)(14,160.97592684906726)(15,163.7897364863335)(16,165.6928815515041)(17,166.65273064969477)(18,163.7897364863335)(19,160.97592684906726)(20,nan)(21,nan)(22,nan)(23,nan)(24,188.14500086724917)(25,177.5852333780714)(26,172.52972650401387)(27,155.49249582586248)(28,172.52972650401387)(29,182.78887791090543)(30,183.84776310850236)(31,184.9127823656466)(32,187.0613654014066)(33,188.14500086724917)(34,188.14500086724917)(35,nan)(36,nan)(37,nan)(38,179.64867439581087)(39,168.58914210463126)(40,173.52918104210445)(41,177.5852333780714)(42,177.5852333780714)(43,174.53442536143424)(44,165.6928815515041)(45,180.68936857950385)(46,178.6139741695516)(47,169.56576906952336)(48,165.6928815515041)(49,162.84637702113662)(50,160.97592684906726)(51,160.04877373483913)(52,154.59692495035372)(53,161.90845090664698)(54,162.84637702113662)(55,159.12696062960518)(56,158.21045677717683)(57,nan)(58,nan)(59,151.06592630451914)(60,142.58724736030138)(61,136.93693467412865)(62,133.03860595958074)(63,135.36407982952485)(64,133.80929076365396)(65,132.2723599733492)(66,129.25125509628117)(67,nan)(68,nan)(69,nan)(70,nan)(71,nan)(72,nan)(73,nan)(74,nan)(75,nan)(76,nan)(77,nan)(78,nan)(79,nan)(80,nan)};
\addlegendentry{6};

\addplot[
    color=iyellow,
    ]
coordinates {
(0,nan)(1,nan)(2,nan)(3,nan)(4,nan)(5,nan)(6,nan)(7,nan)(8,nan)(9,nan)(10,nan)(11,nan)(12,168.58914210463126)(13,169.56576906952336)(14,171.53602840047625)(15,173.52918104210445)(16,175.5454930018647)(17,175.5454930018647)(18,173.52918104210445)(19,169.56576906952336)(20,nan)(21,nan)(22,nan)(23,nan)(24,197.04315364635175)(25,189.23491376948436)(26,182.78887791090543)(27,163.7897364863335)(28,184.9127823656466)(29,193.6580702059513)(30,194.7799199939686)(31,195.90826859169485)(32,199.33268480704808)(33,200.48740730303228)(34,201.6488190383855)(35,nan)(36,nan)(37,197.04315364635175)(38,188.14500086724917)(39,178.6139741695516)(40,180.68936857950385)(41,188.14500086724917)(42,188.14500086724917)(43,184.9127823656466)(44,168.58914210463126)(45,188.14500086724917)(46,184.9127823656466)(47,180.68936857950385)(48,176.56241769755178)(49,173.52918104210445)(50,170.54805357686834)(51,170.54805357686834)(52,169.56576906952336)(53,172.52972650401387)(54,172.52972650401387)(55,169.56576906952336)(56,168.58914210463126)(57,nan)(58,nan)(59,156.39325469067515)(60,151.9410423129619)(61,145.9200662802185)(62,145.9200662802185)(63,145.07962854192903)(64,141.7660052464835)(65,140.13768379557115)(66,138.52806521170294)(67,nan)(68,nan)(69,nan)(70,nan)(71,nan)(72,nan)(73,nan)(74,nan)(75,nan)(76,nan)(77,nan)(78,nan)(79,nan)(80,nan)};
\addlegendentry{4};

\addplot[
    color=igreen,
    ]
coordinates {
(0,nan)(1,nan)(2,nan)(3,nan)(4,nan)(5,nan)(6,nan)(7,nan)(8,nan)(9,nan)(10,nan)(11,nan)(12,179.64867439581087)(13,180.68936857950385)(14,181.73609144327253)(15,182.78887791090543)(16,183.84776310850236)(17,180.68936857950385)(18,179.64867439581087)(19,177.5852333780714)(20,nan)(21,nan)(22,nan)(23,nan)(24,209.96928336801093)(25,201.6488190383855)(26,195.90826859169485)(27,176.56241769755178)(28,184.9127823656466)(29,200.48740730303228)(30,205.1735783080468)(31,207.55758045272898)(32,209.96928336801093)(33,212.40900892134297)(34,211.18562305262122)(35,nan)(36,nan)(37,213.63948179230545)(38,199.33268480704808)(39,189.23491376948436)(40,192.54268179725335)(41,199.33268480704808)(42,199.33268480704808)(43,197.04315364635175)(44,178.6139741695516)(45,195.90826859169485)(46,193.6580702059513)(47,190.3311404729313)(48,187.0613654014066)(49,183.84776310850236)(50,180.68936857950385)(51,179.64867439581087)(52,176.56241769755178)(53,181.73609144327253)(54,183.84776310850236)(55,179.64867439581087)(56,nan)(57,nan)(58,nan)(59,169.56576906952336)(60,161.90845090664698)(61,158.21045677717683)(62,160.04877373483913)(63,160.97592684906726)(64,156.39325469067515)(65,151.9410423129619)(66,nan)(67,nan)(68,nan)(69,nan)(70,nan)(71,nan)(72,nan)(73,nan)(74,nan)(75,nan)(76,nan)(77,nan)(78,nan)(79,nan)(80,nan)};
\addlegendentry{2};
\addplot[
    color=black,
    ]
    coordinates {
(0,nan)(1,nan)(2,nan)(3,nan)(4,nan)(5,nan)(6,nan)(7,nan)(8,nan)(9,nan)(10,nan)(11,nan)(12,190.3311404729313)(13,191.43371755306833)(14,192.54268179725335)(15,194.7799199939686)(16,198.18461302324906)(17,198.18461302324906)(18,195.90826859169485)(19,189.23491376948436)(20,nan)(21,nan)(22,nan)(23,nan)(24,219.8995964441766)(25,212.40900892134297)(26,206.36213675586595)(27,184.9127823656466)(28,198.18461302324906)(29,214.87708272009507)(30,212.40900892134297)(31,218.63306796596578)(32,223.74337368352585)(33,225.0395058615972)(34,225.0395058615972)(35,nan)(36,nan)(37,226.34314645699226)(38,213.63948179230545)(39,200.48740730303228)(40,203.99186545315516)(41,212.40900892134297)(42,212.40900892134297)(43,211.18562305262122)(44,189.23491376948436)(45,216.12185299712482)(46,205.1735783080468)(47,201.6488190383855)(48,198.18461302324906)(49,194.7799199939686)(50,192.54268179725335)(51,190.3311404729313)(52,191.43371755306833)(53,193.6580702059513)(54,194.7799199939686)(55,190.3311404729313)(56,180.68936857950385)(57,181.73609144327253)(58,181.73609144327253)(59,180.68936857950385)(60,170.54805357686834)(61,162.84637702113662)(62,159.12696062960518)(63,160.97592684906726)(64,159.12696062960518)(65,156.39325469067515)(66,155.49249582586248)(67,nan)(68,nan)(69,nan)(70,nan)(71,nan)(72,nan)(73,nan)(74,nan)(75,nan)(76,nan)(77,nan)(78,nan)(79,nan)(80,nan)
    };
    \addlegendentry{0}
    
\addplot[
    color=icyan,
    ]
coordinates {
(0,nan)(1,nan)(2,nan)(3,nan)(4,nan)(5,nan)(6,nan)(7,nan)(8,nan)(9,nan)(10,nan)(11,nan)(12,200.48740730303228)(13,201.6488190383855)(14,202.8169587634771)(15,203.99186545315516)(16,201.6488190383855)(17,203.99186545315516)(18,206.36213675586595)(19,197.04315364635175)(20,nan)(21,nan)(22,nan)(23,nan)(24,234.3251202788159)(25,225.0395058615972)(26,218.63306796596578)(27,189.23491376948436)(28,203.99186545315516)(29,227.65433896553014)(30,230.29955496661373)(31,231.63366671648822)(32,235.68255189130872)(33,238.4210512332145)(34,237.04784702513638)(35,nan)(36,nan)(37,246.828951447135)(38,226.34314645699226)(39,213.63948179230545)(40,217.37383415501247)(41,222.45470667747657)(42,222.45470667747657)(43,221.17346184722248)(44,197.04315364635175)(45,223.74337368352585)(46,222.45470667747657)(47,216.12185299712482)(48,211.18562305262122)(49,206.36213675586595)(50,205.1735783080468)(51,203.99186545315516)(52,203.99186545315516)(53,206.36213675586595)(54,207.55758045272898)(55,203.99186545315516)(56,193.6580702059513)(57,nan)(58,nan)(59,189.23491376948436)(60,182.78887791090543)(61,174.53442536143424)(62,174.53442536143424)(63,172.52972650401387)(64,168.58914210463126)(65,165.6928815515041)(66,nan)(67,nan)(68,nan)(69,nan)(70,nan)(71,nan)(72,nan)(73,nan)(74,nan)(75,nan)(76,nan)(77,nan)(78,nan)(79,nan)(80,nan)};
\addlegendentry{-2};

\addplot[
    color=iblue,
    ]
coordinates {
(0,nan)(1,nan)(2,nan)(3,nan)(4,nan)(5,nan)(6,nan)(7,nan)(8,nan)(9,nan)(10,nan)(11,nan)(12,213.63948179230545)(13,214.87708272009507)(14,217.37383415501247)(15,219.8995964441766)(16,221.17346184722248)(17,222.45470667747657)(18,219.8995964441766)(19,216.12185299712482)(20,nan)(21,nan)(22,nan)(23,nan)(24,261.5061646776419)(25,238.4210512332145)(26,231.63366671648822)(27,207.55758045272898)(28,207.55758045272898)(29,245.40732129724032)(30,243.99387913449723)(31,246.828951447135)(32,249.6969657132611)(33,252.59830469993753)(34,252.59830469993753)(35,nan)(36,nan)(37,nan)(38,238.4210512332145)(39,226.34314645699226)(40,230.29955496661373)(41,237.04784702513638)(42,237.04784702513638)(43,234.3251202788159)(44,195.90826859169485)(45,237.04784702513638)(46,230.29955496661373)(47,225.0395058615972)(48,222.45470667747657)(49,217.37383415501247)(50,213.63948179230545)(51,213.63948179230545)(52,216.12185299712482)(53,219.8995964441766)(54,217.37383415501247)(55,212.40900892134297)(56,201.6488190383855)(57,nan)(58,nan)(59,200.48740730303228)(60,190.3311404729313)(61,183.84776310850236)(62,177.5852333780714)(63,181.73609144327253)(64,178.6139741695516)(65,175.5454930018647)(66,nan)(67,nan)(68,nan)(69,nan)(70,nan)(71,nan)(72,nan)(73,nan)(74,nan)(75,nan)(76,nan)(77,nan)(78,nan)(79,nan)(80,nan)};
\addlegendentry{-4};

\addplot[
    color=iindigo,
    ]
coordinates {
(0,nan)(1,nan)(2,nan)(3,nan)(4,nan)(5,nan)(6,nan)(7,nan)(8,nan)(9,nan)(10,nan)(11,nan)(12,225.0395058615972)(13,226.34314645699226)(14,228.9731271349986)(15,231.63366671648822)(16,235.68255189130872)(17,235.68255189130872)(18,231.63366671648822)(19,223.74337368352585)(20,nan)(21,nan)(22,nan)(23,nan)(24,277.05613042340593)(25,252.59830469993753)(26,245.40732129724032)(27,217.37383415501247)(28,254.06159179290393)(29,260.0)(30,258.50251019256234)(31,261.5061646776419)(32,264.5447199466984)(33,267.61858152730787)(34,267.61858152730787)(35,nan)(36,nan)(37,nan)(38,254.06159179290393)(39,238.4210512332145)(40,242.58857779956992)(41,251.14344552045984)(42,251.14344552045984)(43,248.25881701670832)(44,216.12185299712482)(45,251.14344552045984)(46,245.40732129724032)(47,239.8022103323442)(48,235.68255189130872)(49,231.63366671648822)(50,228.9731271349986)(51,226.34314645699226)(52,225.0395058615972)(53,230.29955496661373)(54,231.63366671648822)(55,226.34314645699226)(56,211.18562305262122)(57,nan)(58,nan)(59,212.40900892134297)(60,202.8169587634771)(61,194.7799199939686)(62,185.98397121658357)(63,191.43371755306833)(64,189.23491376948436)(65,185.98397121658357)(66,183.84776310850236)(67,nan)(68,nan)(69,nan)(70,nan)(71,nan)(72,nan)(73,nan)(74,nan)(75,nan)(76,nan)(77,nan)(78,nan)(79,nan)(80,nan)};
\addlegendentry{-6};

\addplot[
    color=ipurple,
    ]
coordinates {
(0,nan)(1,nan)(2,nan)(3,nan)(4,nan)(5,nan)(6,nan)(7,nan)(8,nan)(9,nan)(10,nan)(11,nan)(12,239.8022103323442)(13,239.8022103323442)(14,242.58857779956992)(15,246.828951447135)(16,251.14344552045984)(17,251.14344552045984)(18,246.828951447135)(19,239.8022103323442)(20,nan)(21,nan)(22,nan)(23,nan)(24,290.159257083858)(25,267.61858152730787)(26,261.5061646776419)(27,232.97550689710764)(28,255.5333556217653)(29,273.87386934825736)(30,273.87386934825736)(31,277.05613042340593)(32,281.8989862783688)(33,283.532010492967)(34,283.532010492967)(35,nan)(36,nan)(37,281.8989862783688)(38,267.61858152730787)(39,252.59830469993753)(40,257.013645291753)(41,267.61858152730787)(42,266.0772119191614)(43,261.5061646776419)(44,238.4210512332145)(45,266.0772119191614)(46,258.50251019256234)(47,252.59830469993753)(48,249.6969657132611)(49,245.40732129724032)(50,239.8022103323442)(51,237.04784702513638)(52,235.68255189130872)(53,243.99387913449723)(54,245.40732129724032)(55,239.8022103323442)(56,222.45470667747657)(57,nan)(58,nan)(59,222.45470667747657)(60,214.87708272009507)(61,205.1735783080468)(62,199.33268480704808)(63,203.99186545315516)(64,201.6488190383855)(65,198.18461302324906)(66,194.7799199939686)(67,nan)(68,nan)(69,nan)(70,nan)(71,nan)(72,nan)(73,nan)(74,nan)(75,nan)(76,nan)(77,nan)(78,nan)(79,nan)(80,nan)};
\addlegendentry{-8};
\end{axis}
\end{tikzpicture}

%% file: figure/pitch_variaiton.tex
\begin{tikzpicture}

\global\definecolor{ired}{RGB}{255,49,48}
\global\definecolor{iorange}{RGB}{255,149,0}
\global\definecolor{iyellow}{RGB}{255,204,0}
\global\definecolor{igreen}{RGB}{52,199,89}
\global\definecolor{icyan}{RGB}{50,173,230}
\global\definecolor{iblue}{RGB}{0,122,255}
\global\definecolor{iindigo}{RGB}{88,86,214}
\global\definecolor{ipurple}{RGB}{175,82,222}

\begin{axis}[
    unbounded coords=jump,
    xlabel={Time Frame},
    ylabel={$f_0$\phantom{z}\\(Hz)},
    ymode=log,  
    log ticks with fixed point,
    width=\linewidth,
    height=\linewidth*0.45,
    xmin=5, xmax=65,
    ymin=160, ymax=270,
    xtick={5,25,45,65},
    xticklabels={0,20,40,60},
    ytick={160,200,240},
    ymajorgrids=true,
    grid style=dashed,
    label style={inner sep=0pt},
    every axis y label/.style={
        at={(0,0.8)},
        anchor=south east,
        align=center,
        outer sep=0pt,
    },
    outer sep=0pt,
]

\addplot[
    color=ired,
    ]
coordinates {
(0,nan)(1,nan)(2,nan)(3,nan)(4,nan)(5,nan)(6,nan)(7,nan)(8,nan)(9,219.8995964441766)(10,219.8995964441766)(11,218.63306796596578)(12,218.63306796596578)(13,218.63306796596578)(14,222.45470667747657)(15,219.8995964441766)(16,nan)(17,nan)(18,nan)(19,nan)(20,nan)(21,227.65433896553014)(22,222.45470667747657)(23,216.12185299712482)(24,212.40900892134297)(25,208.75994928447787)(26,208.75994928447787)(27,209.96928336801093)(28,211.18562305262122)(29,213.63948179230545)(30,216.12185299712482)(31,nan)(32,nan)(33,nan)(34,202.8169587634771)(35,197.04315364635175)(36,191.43371755306833)(37,185.98397121658357)(38,183.84776310850236)(39,180.68936857950385)(40,188.14500086724917)(41,191.43371755306833)(42,192.54268179725335)(43,194.7799199939686)(44,195.90826859169485)(45,193.6580702059513)(46,192.54268179725335)(47,192.54268179725335)(48,nan)(49,nan)(50,nan)(51,nan)(52,nan)(53,195.90826859169485)(54,199.33268480704808)(55,202.8169587634771)(56,200.48740730303228)(57,nan)(58,nan)(59,nan)(60,nan)};
\addlegendentry{0};
\addplot[
    color=iorange,
    ]
coordinates {
(0,nan)(1,nan)(2,nan)(3,nan)(4,nan)(5,nan)(6,nan)(7,nan)(8,nan)(9,nan)(10,nan)(11,nan)(12,nan)(13,227.65433896553014)(14,226.34314645699226)(15,225.0395058615972)(16,230.29955496661373)(17,239.8022103323442)(18,243.99387913449723)(19,243.99387913449723)(20,232.97550689710764)(21,nan)(22,nan)(23,nan)(24,nan)(25,226.34314645699226)(26,219.8995964441766)(27,211.18562305262122)(28,202.8169587634771)(29,200.48740730303228)(30,199.33268480704808)(31,200.48740730303228)(32,201.6488190383855)(33,203.99186545315516)(34,205.1735783080468)(35,205.1735783080468)(36,206.36213675586595)(37,211.18562305262122)(38,nan)(39,nan)(40,nan)(41,nan)(42,214.87708272009507)(43,207.55758045272898)(44,201.6488190383855)(45,201.6488190383855)(46,199.33268480704808)(47,194.7799199939686)(48,192.54268179725335)(49,191.43371755306833)(50,192.54268179725335)(51,193.6580702059513)(52,194.7799199939686)(53,192.54268179725335)(54,188.14500086724917)(55,182.78887791090543)(56,178.6139741695516)(57,177.5852333780714)(58,179.64867439581087)(59,180.68936857950385)(60,180.68936857950385)(61,187.0613654014066)(62,nan)(63,nan)(64,nan)(65,nan)};
\addlegendentry{1};
\addplot[
    color=iyellow,
    ]
coordinates {
(0,nan)(1,nan)(2,nan)(3,nan)(4,nan)(5,nan)(6,nan)(7,nan)(8,nan)(9,nan)(10,223.74337368352585)(11,223.74337368352585)(12,222.45470667747657)(13,222.45470667747657)(14,222.45470667747657)(15,225.0395058615972)(16,225.0395058615972)(17,225.0395058615972)(18,225.0395058615972)(19,nan)(20,nan)(21,nan)(22,216.12185299712482)(23,212.40900892134297)(24,208.75994928447787)(25,207.55758045272898)(26,205.1735783080468)(27,201.6488190383855)(28,198.18461302324906)(29,198.18461302324906)(30,202.8169587634771)(31,207.55758045272898)(32,209.96928336801093)(33,212.40900892134297)(34,214.87708272009507)(35,nan)(36,nan)(37,nan)(38,nan)(39,195.90826859169485)(40,194.7799199939686)(41,194.7799199939686)(42,194.7799199939686)(43,197.04315364635175)(44,197.04315364635175)(45,195.90826859169485)(46,195.90826859169485)(47,193.6580702059513)(48,192.54268179725335)(49,192.54268179725335)(50,194.7799199939686)(51,197.04315364635175)(52,198.18461302324906)(53,199.33268480704808)(54,193.6580702059513)(55,193.6580702059513)(56,nan)(57,nan)(58,nan)(59,nan)(60,nan)};
\addlegendentry{2};
\addplot[
    color=igreen,
    ]
coordinates {
(0,nan)(1,nan)(2,nan)(3,nan)(4,nan)(5,nan)(6,nan)(7,nan)(8,nan)(9,nan)(10,228.9731271349986)(11,222.45470667747657)(12,219.8995964441766)(13,216.12185299712482)(14,212.40900892134297)(15,214.87708272009507)(16,214.87708272009507)(17,212.40900892134297)(18,nan)(19,nan)(20,nan)(21,232.97550689710764)(22,222.45470667747657)(23,217.37383415501247)(24,213.63948179230545)(25,213.63948179230545)(26,213.63948179230545)(27,208.75994928447787)(28,203.99186545315516)(29,202.8169587634771)(30,202.8169587634771)(31,203.99186545315516)(32,202.8169587634771)(33,nan)(34,nan)(35,nan)(36,nan)(37,nan)(38,195.90826859169485)(39,194.7799199939686)(40,191.43371755306833)(41,190.3311404729313)(42,190.3311404729313)(43,190.3311404729313)(44,190.3311404729313)(45,191.43371755306833)(46,191.43371755306833)(47,190.3311404729313)(48,188.14500086724917)(49,184.9127823656466)(50,182.78887791090543)(51,181.73609144327253)(52,183.84776310850236)(53,188.14500086724917)(54,194.7799199939686)(55,195.90826859169485)(56,185.98397121658357)(57,nan)(58,nan)(59,nan)(60,nan)(61,nan)(62,nan)};
\addlegendentry{3};
\addplot[
    color=icyan,
    ]
coordinates {
(0,nan)(1,nan)(2,nan)(3,nan)(4,nan)(5,nan)(6,nan)(7,nan)(8,226.34314645699226)(9,223.74337368352585)(10,222.45470667747657)(11,219.8995964441766)(12,217.37383415501247)(13,216.12185299712482)(14,nan)(15,nan)(16,nan)(17,nan)(18,nan)(19,225.0395058615972)(20,218.63306796596578)(21,213.63948179230545)(22,211.18562305262122)(23,212.40900892134297)(24,214.87708272009507)(25,214.87708272009507)(26,212.40900892134297)(27,209.96928336801093)(28,208.75994928447787)(29,202.8169587634771)(30,nan)(31,nan)(32,nan)(33,nan)(34,201.6488190383855)(35,197.04315364635175)(36,191.43371755306833)(37,190.3311404729313)(38,188.14500086724917)(39,189.23491376948436)(40,193.6580702059513)(41,192.54268179725335)(42,190.3311404729313)(43,189.23491376948436)(44,189.23491376948436)(45,188.14500086724917)(46,185.98397121658357)(47,182.78887791090543)(48,181.73609144327253)(49,180.68936857950385)(50,182.78887791090543)(51,189.23491376948436)(52,190.3311404729313)(53,189.23491376948436)(54,190.3311404729313)(55,193.6580702059513)(56,nan)(57,nan)(58,nan)(59,nan)};
\addlegendentry{4};
\addplot[
    color=iblue,
    ]
coordinates {
(0,nan)(1,nan)(2,nan)(3,nan)(4,nan)(5,nan)(6,nan)(7,nan)(8,nan)(9,217.37383415501247)(10,218.63306796596578)(11,219.8995964441766)(12,222.45470667747657)(13,230.29955496661373)(14,239.8022103323442)(15,242.58857779956992)(16,241.19137040474044)(17,nan)(18,nan)(19,nan)(20,nan)(21,nan)(22,225.0395058615972)(23,219.8995964441766)(24,217.37383415501247)(25,216.12185299712482)(26,217.37383415501247)(27,221.17346184722248)(28,221.17346184722248)(29,218.63306796596578)(30,217.37383415501247)(31,214.87708272009507)(32,212.40900892134297)(33,209.96928336801093)(34,208.75994928447787)(35,205.1735783080468)(36,nan)(37,nan)(38,nan)(39,nan)(40,nan)(41,199.33268480704808)(42,189.23491376948436)(43,185.98397121658357)(44,184.9127823656466)(45,183.84776310850236)(46,181.73609144327253)(47,180.68936857950385)(48,180.68936857950385)(49,182.78887791090543)(50,184.9127823656466)(51,185.98397121658357)(52,185.98397121658357)(53,183.84776310850236)(54,181.73609144327253)(55,180.68936857950385)(56,181.73609144327253)(57,183.84776310850236)(58,184.9127823656466)(59,189.23491376948436)(60,189.23491376948436)(61,nan)(62,nan)(63,nan)(64,nan)};
\addlegendentry{5};
\addplot[
    color=iindigo,
    ]
coordinates {
(0,nan)(1,nan)(2,nan)(3,nan)(4,nan)(5,nan)(6,nan)(7,nan)(8,nan)(9,nan)(10,nan)(11,235.68255189130872)(12,235.68255189130872)(13,234.3251202788159)(14,232.97550689710764)(15,239.8022103323442)(16,241.19137040474044)(17,241.19137040474044)(18,nan)(19,nan)(20,nan)(21,nan)(22,nan)(23,230.29955496661373)(24,225.0395058615972)(25,218.63306796596578)(26,212.40900892134297)(27,209.96928336801093)(28,208.75994928447787)(29,209.96928336801093)(30,209.96928336801093)(31,207.55758045272898)(32,205.1735783080468)(33,202.8169587634771)(34,201.6488190383855)(35,201.6488190383855)(36,200.48740730303228)(37,195.90826859169485)(38,nan)(39,nan)(40,nan)(41,nan)(42,nan)(43,192.54268179725335)(44,185.98397121658357)(45,183.84776310850236)(46,182.78887791090543)(47,184.9127823656466)(48,188.14500086724917)(49,190.3311404729313)(50,193.6580702059513)(51,195.90826859169485)(52,193.6580702059513)(53,189.23491376948436)(54,184.9127823656466)(55,178.6139741695516)(56,179.64867439581087)(57,183.84776310850236)(58,185.98397121658357)(59,190.3311404729313)(60,191.43371755306833)(61,nan)(62,nan)(63,nan)(64,nan)(65,nan)};
\addlegendentry{6};
\addplot[
    color=ipurple,
    ]
coordinates {
(0,nan)(1,nan)(2,nan)(3,nan)(4,nan)(5,nan)(6,nan)(7,nan)(8,nan)(9,nan)(10,227.65433896553014)(11,227.65433896553014)(12,226.34314645699226)(13,221.17346184722248)(14,221.17346184722248)(15,222.45470667747657)(16,223.74337368352585)(17,222.45470667747657)(18,nan)(19,nan)(20,nan)(21,nan)(22,nan)(23,nan)(24,217.37383415501247)(25,213.63948179230545)(26,208.75994928447787)(27,206.36213675586595)(28,205.1735783080468)(29,206.36213675586595)(30,209.96928336801093)(31,207.55758045272898)(32,202.8169587634771)(33,202.8169587634771)(34,203.99186545315516)(35,205.1735783080468)(36,nan)(37,nan)(38,nan)(39,nan)(40,nan)(41,198.18461302324906)(42,194.7799199939686)(43,192.54268179725335)(44,194.7799199939686)(45,195.90826859169485)(46,193.6580702059513)(47,193.6580702059513)(48,194.7799199939686)(49,197.04315364635175)(50,199.33268480704808)(51,198.18461302324906)(52,195.90826859169485)(53,nan)(54,nan)(55,nan)(56,nan)(57,193.6580702059513)(58,192.54268179725335)(59,191.43371755306833)(60,191.43371755306833)(61,201.6488190383855)(62,nan)(63,nan)(64,nan)(65,nan)(66,nan)};
\addlegendentry{7};

\addplot[
    color=white,
    ]
coordinates {(1,250)};
\addlegendentry{\phantom{----0}};
\legend{}; 

\end{axis}
\end{tikzpicture}

%% file: paper/discussion.tex
\vspace{-0.5\baselineskip}
\section{Conclusion}
\vspace{-0.5\baselineskip}
In this paper, we propose PITS, an end-to-end variational pitch inference TTS. 
With our introduced method, the Yingram encoder, the Yingram decoding loss, and adversarial pitch-shifted training could build a pitch-controllable TTS model without directly modeling fundamental frequency, which is not well-defined in some cases. 
We also introduce Q-VAE with single-stage training, however, it reduces speech quality and intelligibility significantly.
With our experiments, we could find the best architecture and losses, PITS (A+D), which could generate high naturalness and intelligibility as ground truth speeches.
In addition, PITS could generate pitch-shifted speech by scope shift of latent variables without quality degradation.

%% file: paper/appendix.tex
\section{Voice Conversion} \label{appx:vc}
We undertake voice conversion (VC) using the pretrained PITS model, since our model disentangles linguistic and pitch information following the principles in VC studies.
We perform an any-to-many voice conversion, which aims to convert the speech of an unseen source speaker to the speech of a specific seen speaker while preserving the linguistic and pitch information of the source speech.  
To better preserve the linguistic information of speech, we utilized both voice and text as inputs.
However, different from the seen speakers, we do not have the speaker embedding of the unseen source speakers, which is required by modules such as the posterior encoders and the flow. 
Therefore, we optimize the speaker embedding of the source speaker $e_s$ by minimizing the KL divergence from VITS \cite{vits} as:
\newcommand{\argmin}{\mathop{\mathrm{argmin}}}
\begin{equation} \label{eq:e_s}
    e_s = \argmin_{e_s}  \log q_\phi (z|S, Y, e_s) -  \log p_\theta(z|c_\mathrm{text}, A)
\end{equation}
where $S$ and $Y$ represent linear spectrogram and Yingram of source speech, respectively, $c_\mathrm{text}$ is a phonemes of input speech, $A$ is a optimal alignment by MAS.
Furthermore, this process enables the acquisition of the optimal alignment $A$ between text and speech, which is utilized to expand the prior distribution $\mu$ and $\sigma$ from the text encoder to $\mu_A$ and $\sigma_A$.
We then obtain $z_p$ from $\mathcal{N}(z_p;\mu_A, \sigma_A)$, which is subsequently fed into the inverse flow module $f^{-1}$.
Empirically, one minute of training for source speaker embedding is sufficient to obtain a high-quality alignment.

To preserve the linguistic information of the text, we obtain $z_{\mathrm{spec}}$ by splitting $z = f^{-1}(z_p, e_t)$ where $e_t$ is the target speaker embedding. 
Furthermore, to maintain the pitch information, we directly utilize $z_\mathrm{yin}$ obtained from the source speech through the Yingram encoder with speaker condition $e_t$. 
Finally, pitch-guided speaker-converted speech is synthesized via the decoder from the concatenated latent $z$, which is composed of $z_{\mathrm{spec}}$ and cropped Yingram latent $z_{\mathrm{crop}}^{\mathrm{shift}} = \mathrm{shift\_crop}(z_\mathrm{yin}, s)$ with scope shift $s$ where $\mathrm{shift\_crop}(z, s) = z[:, 15+s:65+s]$.

Direct utilization of $z_\mathrm{crop}^{\mathrm{shift}}$ from the unseen source speaker preserved pitch information, but degraded the audio quality of the converted speech. 
We first tried to optimize $z_\mathrm{crop}^{\mathrm{shift}}$ by applying an exponential regularization and $L_1$ loss between the decoded Yingram and the cropped Yingram $Y_{\mathrm{crop}}^{\mathrm{shift}}$ of the source speech $y$ with scope shift $s$.
\begin{align*}
    \mathcal{L}_{\mathrm{yin}}^{\mathrm{VC}} = \left\Vert Y_{\mathrm{crop}}^{\mathrm{shift}} - \mathrm{Dec}_\mathrm{yin} \left(z_\mathrm{crop}^{\mathrm{shift}} \right)\right\Vert_{1} + \lambda \left\Vert\exp \left(z_\mathrm{crop}^{\mathrm{shift}}\right)\right\Vert_{1}
\end{align*}
While the quality of the generated audio is improved through the above modification, it still failed to achieve the quality of speech generated by PITS for TTS purpose. 

The iterative synthesis illustrated in Algorithm~\ref{alg:vc} generates more natural results. 
In each $i$th iteration, the $i$th generated audio $y_i$ is produced by feeding the  $z_{\mathrm{crop}_i}$, cropped Yingram latent of the preceding audio output $y_{i-1}$, instead of $z_\mathrm{crop}^{\mathrm{shift}}$ when creating the concatenated latent $z_i$. However, substituting $z_{\mathrm{spec}}$ in the $i$th iteration with the output $\mathrm{Enc}_\mathrm{spec}(\mathcal{S}(y_i), e_t)$ of the STFT encoder led to the vanishing of linguistic information.
Therefore, we used the same $z_{\mathrm{spec}}$ from the $0$th iteration for every iteration to preserve linguistic information.
Three iterations are sufficient for generating high-quality, pitch-conserving speech.

With pretrained PITS, it has been observed that when the pitch of the source speech falls outside the pitch range of the target speaker, the generated speech occasionally exhibits pitch deviations from the source speech by multiple octaves.
To address this issue, we conducted finetuning using the NUS-48E singing voice dataset \cite{nus48e} with the original PITS objective.
This resulted in an improvement in the expressiveness and expanded pitch range capabilities of the model, resulting in a reduction in the occurrence of octave-scale pitch differences between the source and generated speech.
In addition, shifting $z_\mathrm{yin}$ to match each speaker's pitch range effectively mitigates octave-scale pitch differences and prevents quality degradation.

%
%

\begin{algorithm}[h]
   \caption{Voice Conversion by PITS}
   \label{alg:vc}
\begin{algorithmic}
    \STATE {\bfseries Input:} source speech $y$, phoneme $c_\mathrm{text}$, target speaker embedding $e_t$, scope shift $s$, number of iteration $N$
    \STATE {\bfseries Module:} Yingram operator $\mathcal{Y}$, spectrogram operator $\mathcal{S}$, prior encoder $\mathrm{Enc}_\mathrm{p}$, 
           Yingram encoder $\mathrm{Enc}_\mathrm{yin}$, STFT encoder $\mathrm{Enc}_\mathrm{spec}$, flow $f$, decoder $G$
    \STATE {\bfseries Output:} Converted voice $y_N$
    \vspace{8pt}
    \STATE Calculate $Y \leftarrow \mathcal{Y}(y)$, $S  \leftarrow  \mathcal{S}(y)$
    \STATE Calculate prior distribution $(\mu, \sigma) \leftarrow \mathrm{Enc}_\mathrm{p}(c_\mathrm{text}) $ 
    \STATE Initialize source speaker embedding $e_s  \sim \mathcal{N}(0,I)$.
    \REPEAT
        \STATE Calculate optimal alignment $A$ by MAS
        \STATE Update $e_s$ through SGD with KL-divergence objective\vspace{-8pt}\[\log q_\phi (z|S, Y, e_s) -  \log p_\theta(z|c_\mathrm{text}, A)\]\vspace{-13pt}
    \UNTIL{$e_s$ is converged}
    \STATE Expand $\mu$ and $\sigma$ with alignment $A$ resulting in $\mu_A$ and $\sigma_A$, respectively.
    \STATE Sample $z_p \sim \mathcal{N}(z_p; \mu_A, \sigma_A)$
    \STATE $z \leftarrow f^{-1}(z_p, e_t)$
    \STATE $z_{\mathrm{spec}} \leftarrow \mathrm{split}(z)[0]$
    \STATE $z_{\mathrm{crop}}^{\mathrm{shift}} \leftarrow \mathrm{shift\_crop}(\mathrm{Enc}_\mathrm{yin}(Y, e_t), s)$
    \STATE $z_{0} \leftarrow [z_{\mathrm{spec}}, z_{\mathrm{crop}}^{\mathrm{shift}}]$
    \STATE $y_{0} \leftarrow G(z_{0}, e_t)$
    \FOR{$i=1$ {\bfseries to} $N$}
        \STATE $z_{\mathrm{crop}_i} \leftarrow \mathrm{shift\_crop}(\mathrm{Enc}_\mathrm{yin}(\mathcal{Y}(y_{i-1}), e_t), 0)$
        \STATE $z_i \leftarrow [z_{\mathrm{spec}}, z_{\mathrm{crop}_i}]$
        \STATE $y_i \leftarrow G(z_i, e_t)$
    \ENDFOR
    \STATE \textbf{return} $y_N$
\end{algorithmic}
\end{algorithm}